\documentclass[cernpreprint, atlasdraft=false, UKenglish, texmf, orcidlogo]{atlasdoc}
 
\usepackage{atlaspackage}
\usepackage{atlasbiblatex}
 
\usepackage{atlasphysics}

\graphicspath{{logos/}{figures/}{figures/AUX/}}
 
\usepackage{ANA-EXOT-2022-37-PAPER-defs}
\usepackage{siunitx}

\AtlasTitle{Search for non-resonant production of semi-visible jets using Run~2 data in ATLAS}

\AtlasAbstract{
Semi-visible jets, with
a significant contribution to the event's missing transverse momentum,
can arise in strongly interacting dark sectors.
This results in an event topology
where one of the jets can be aligned with the direction of the missing
transverse momentum.
The first search for semi-visible jets produced
via a $t$-channel mediator exchange is presented.
The analysis uses proton-proton collisions with an integrated
luminosity of 139~fb$^{-1}$ and a centre-of-mass energy of 13~TeV,
collected  with the ATLAS detector during the Run~2 of the LHC.
No excess over Standard Model predictions is observed.
Assuming a coupling strength of unity between the mediator, a Standard Model quark and a dark quark,
mediator masses up to 2.7~TeV are excluded at the 95\% confidence level.
Upper limits on the coupling strength are also derived.
}

\AtlasRefCode{EXOT-2022-37}
 
\PreprintIdNumber{CERN-EP-2023-084}

\AtlasJournal{Phys.\ Lett.\ B.}

\AtlasJournalRef{Phys. Lett. B 848 (2024) 138324}
\AtlasDOI{10.1016/j.physletb.2023.138324}

\hypersetup{pdftitle={ATLAS document},pdfauthor={The ATLAS Collaboration}}
 
\begin{document}
 
\maketitle

\section{Introduction}
\label{sec:intro}

Collider searches for dark matter (DM) have often focussed on scenarios
in which DM particles are produced in association with heavy Standard Model (SM) particles
or photons or jets.
However, no experimental evidence confirming the existence of DM has been found so far.
Several proposed models~\cite{Strassler:2006im, Strassler:2006ri, Han:2007ae, Schwaller:2015gea}
include a strongly coupled dark sector, giving
rise to unexplored collider-event topologies, and thus motivating the present search.
 
A feature of the strongly coupled dark sector is the so-called \enquote{dark shower} (DS), emulating the QCD parton shower.
Since the DS contains families of dark quarks which bind into dark
mesons at energies lower than the dark confinement scale,
the subsequent hadronisation can give rise to
flavour-diagonal and off-diagonal $\pi_{\mathrm{d}}$ and $\rho_{\mathrm{d}}$ mesons, with spin~0 and 1 respectively.
Depending on the flavour content of the dark shower, these dark mesons can be stable or can decay back into the SM particles.

The models described above are particularly useful as event-topology generators, and depending on the
fraction of stable dark hadrons among all dark hadrons in the event, denoted as \Rinv, different
experimental signatures can arise. If this fraction is unity, the signature would consist of
missing transverse momentum, the magnitude referred to as \met, produced along with a jet (or jets) from the SM sector, resembling the
monojet signature. On the other hand, if the fraction is zero, the signature would resemble
QCD multijet events with no intrinsic missing transverse momentum.
This signature is often referred to as a \enquote{dark jet}~\cite{Park:2017rfb}.
Semi-visible jet (SVJ)~\cite{Cohen:2015toa,Cohen:2017pzm} signatures arise when the fraction has intermediate
values, resulting in jets geometrically encompassing dark hadrons.

A search for SVJ signatures in the $s$-channel production mode has been presented
by the CMS Collaboration~\cite{CMS-EXO-19-020}, and
excludes mediator masses up to 5.1~\TeV,
depending on other signal parameters.
This Letter presents a first search for $t$-channel production of SVJs
using the full LHC \RunTwo~ dataset of 13~\TeV\ proton--proton collisions with an integrated luminosity of 139~fb$^{-1}$ recorded by the ATLAS experiment.
Searches for $t$-channel production modes probe
a broad class of non-resonant signals.

\begin{figure}[ht]
\centering
\vspace{0.2cm}
\includegraphics[trim=0.0cm 0.3cm 0 0.56cm,width=0.35\textwidth]{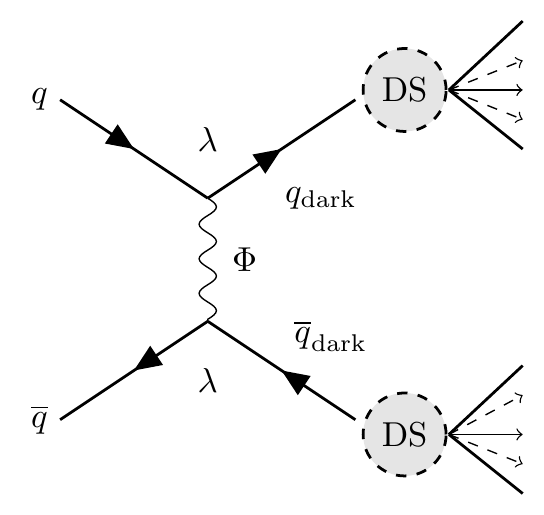}
\caption{A diagram illustrating the production
of semi-visible jets via a $t$-channel mediator, $\Phi$,
producing a pair of dark quarks,
labelled $q_{\mathrm{dark}}$.
DS denotes the dark shower
which produces a final state consisting of SM hadrons and
dark hadrons, governed by the \Rinv\ fraction.
The coupling strength of the $q$--$q_{\mathrm{dark}}$--$\Phi$ interaction is denoted by $\lambda$.
}
\label{fig:model}
\end{figure}

In the $t$-channel production mode, a scalar mediator, $\Phi$,
acts as a portal between the SM and dark sectors.
It couples to a SM quark and a dark quark, and mediates the production of
dark quarks, which then hadronise to stable and unstable dark hadrons
through the DS process described above and shown in Figure~\ref{fig:model}.

At the leading order, the two SVJs are produced back-to-back
and the direction of the missing transverse momentum ($\vec{p}^\text{miss}_\text{T}$)
is aligned with one of the two reconstructed jets.
A boost due to additional jets leads to
signatures with the
$\vec{p}^\text{miss}_\text{T}$ not necessarily pointing in the
direction of one of the two SVJs,
since both of them contribute to the $\vec{p}^\text{miss}_\text{T}$.
Events in which the $\vec{p}^\text{miss}_\text{T}$ is aligned with a jet typically contain
a mismeasured jet, and are usually discarded.
This search, however, considers an inclusive fiducial phase space and thus is
complementary to existing searches that have focussed on phase-space regions that suppress
background contributions arising from mismeasured jets.

\FloatBarrier

\section{ATLAS detector}
\label{sec:atlasdet}
 
The ATLAS experiment~\cite{PERF-2007-01} at the LHC is a multipurpose particle detector
with a forward--backward symmetric cylindrical geometry and a near \(4\pi\) coverage in
solid angle.\footnote{\AtlasCoordFootnote}
It consists of an inner tracking detector surrounded by a thin superconducting solenoid
providing a \SI{2}{\tesla} axial magnetic field, electromagnetic and hadron calorimeters, and a muon spectrometer.
The inner tracking detector covers the pseudorapidity range \(|\eta| < 2.5\).
It consists of silicon pixel, silicon microstrip, and transition radiation tracking detectors.
Lead/liquid-argon (LAr) sampling calorimeters provide electromagnetic (EM) energy measurements
with high granularity.
A steel/scintillator-tile hadron calorimeter covers the central pseudorapidity range (\(|\eta| < 1.7\)).
The endcap and forward regions are instrumented with LAr calorimeters
for both the EM and hadronic energy measurements up to \(|\eta| = 4.9\).
The muon spectrometer surrounds the calorimeters and is based on
three large superconducting air-core toroidal magnets with eight coils each.
The field integral of the toroids ranges between \num{2.0} and \SI{6.0}{\tesla\metre}
across most of the detector.
The muon spectrometer includes a system of precision chambers for tracking and fast detectors for triggering.
A two-level trigger system is used to select events.
The first-level trigger is implemented in hardware and uses a subset of the detector information
to accept events at a rate below \SI{100}{\kHz}.
This is followed by a software-based trigger that
reduces the accepted event rate to \SI{1}{\kHz} on average
depending on the data-taking conditions.
An extensive software suite~\cite{ATL-SOFT-PUB-2021-001} is used in data simulation, in the reconstruction
and analysis of real and simulated data, in detector operations, and in the trigger and data acquisition
systems of the experiment.

\FloatBarrier

\section{Signal and background event simulation}
\label{sec:mc}
 
Signal events were modelled at the leading order (LO) in QCD,
using the \MGNLO~\cite{Alwall:2014hca} event generator
to calculate matrix elements (ME) with up to two extra partons at the leading order.
The \NNPDF[3.0lo]~\cite{Ball:2014uwa} parton distribution function (PDF) set was used.
The mediator mass was varied
within the range 1--5~\TeV\ in 0.5~\TeV\ steps.
The Hidden Valley (HV) module~\cite{Carloni:2011kk} of \PYTHIA[8]~\cite{Sjostrand:2014zea}
was used to shower the ME-level event and produce dark hadrons, using the \NNPDF[2.3lo]~\cite{Ball:2014uwa}
PDF set and
the A14 set of tuned parameters~\cite{ATL-PHYS-PUB-2014-021}.
The MLM~\cite{Mangano:2006rw} jet matching scheme, with
the matching parameter set to 100~\GeV, was employed.
 
A detailed description of the HV module is beyond the scope of this Letter, and the HV
parameter settings adopted here provide a simple way to generate this mostly unexplored collider-event topology.
The mass of the dark quark was set to 10~\GeV, the flavour-diagonal $\pi_{\mathrm{d}}$ and $\rho_{\mathrm{d}}$ meson masses
were set to 10~\GeV\ and 20~\GeV\ respectively, and the off-diagonal $\pi_{\mathrm{d}}$ and $\rho_{\mathrm{d}}$
meson masses were set to 4.99~\GeV\ and 9.99~\GeV\ respectively.
The number of flavours in the HV module was set to unity. This implies that
flavour diagonal $\pi_{\mathrm{d}}$ and $\rho_{\mathrm{d}}$ mesons decay promptly
into off-diagonal stable dark $\pi_{\mathrm{d}}$ and $\rho_{\mathrm{d}}$ pairs, and to
SM quarks, following a five-flavour scheme, due to portal interactions of the mediator which couples the
SM sector to the dark sector~\cite{Bernreuther:2019pfb}.
These choices are based on Refs.~\cite{Cohen:2017pzm, Albouy:2022cin}
and kinematic considerations for enabling the relevant decays.
The general topology of the signal events shows negligible sensitivity to the chosen mass values.

The branching fraction of unstable dark mesons decaying to stable dark mesons is taken as the \Rinv\ fraction  defined above,
and is a free parameter of the model.
The signal samples were generated with \Rinv\ values of 0.1, 0.2, 0.4, 0.6, 0.8, and 0.9 for each simulated mediator mass value.
The \PYTHIA[8] HV $\alpha_{\textrm{dark}}$
coupling was chosen to be running at one-loop, and the dark QCD confinement scale value was set to 6.5~\TeV,
based on Ref.~\cite{Albouy:2022cin}. Another free parameter in the model is the strength of the coupling connecting the SM and DM sectors (shown as $\lambda$ in Figure~\ref{fig:model}).
The nominal samples were generated with $\lambda=1$, although $\lambda$ can be
varied up to $4\pi$~\cite{Cohen:1997rt,Colucci:2018vxz}.
The cross-section scales as $\lambda^4$ without having any impact on the kinematic distributions,
or on the validity of the model, if the mediator mass is 2.5~\TeV\ or higher. At lower
mediator masses, there are stronger contributions from resonant production of the mediator as well.

Simulated event samples were used to describe the main background processes -- $W/Z$+jets, $t\bar{t}$, single top, multijet, and
diboson -- as listed in Table~\ref{tab:samples}.
The  Monte Carlo (MC) generated samples were processed through
a detailed ATLAS detector simulation~\cite{SOFT-2010-01}
based on \GEANT~\cite{Agostinelli:2002hh}.
Additional simulated inelastic proton--proton collisions generated using \PYTHIA[8] with the A3 set of tuned parameters~\cite{ATL-PHYS-PUB-2016-017}
and \NNPDF[2.3lo] PDF set were overlaid on each simulated hard-scatter event to model the effects of additional collisions in
the same or neighbouring bunch crossings (referred to as \enquote{pile-up}).
The simulated events were then reconstructed and analysed using the
same procedure and software as used for data events.

\begin{table}[] \small
\centering
\caption{Summary of generators used to simulate background processes,
along with the PDF, parton shower and tune used.}
\resizebox{\textwidth}{!}{
\def\arraystretch{1.5}
\newcolumntype{L}[1]{>{\raggedright\let\newline\\\arraybackslash}p{#1}}
\begin{tabular}{L{0.11\textwidth}L{0.23\textwidth}L{0.18\textwidth}L{0.19\textwidth}L{0.15\textwidth}L{0.1\textwidth}}
\toprule
Process & Generator & ME order & PDF & Parton shower & Tune \\
\hline
$W/Z+$jets &\SHERPA[2.2.11]~\cite{Bothmann:2019yzt, Schumann:2007mg} & NLO (up to 2 jets) & \NNPDF[3.0nnlo]~\cite{Ball:2014uwa} & \SHERPA \MEPSatNLO & \SHERPA \\
\ttbar & \POWHEGBOX[v2]~\cite{Nason:2004rx,Frixione:2007vw,Alioli:2010xd} & NLO & \NNPDF[3.0nlo]~\cite{Ball:2014uwa} & \PYTHIA[8.230] with \NNPDF[2.3lo] & A14~\cite{ATL-PHYS-PUB-2014-021}  \\
Single top & \POWHEGBOX[v2] & NLO & \NNPDF[3.0nnlo] & \PYTHIA[8.230] with \NNPDF[2.3lo] & A14  \\
Multijet & \Pythia[8.230]~\cite{Sjostrand:2014zea}  & LO & \NNPDF[2.3lo]~\cite{Ball:2014uwa}  & \Pythia[8.230] & A14 \\
Diboson & \SHERPA[2.2.1] & NLO (up to 2 jets) & \NNPDF[3.0nnlo] & \SHERPA \MEPSatNLO & \SHERPA \\
\bottomrule
\end{tabular}
}
\label{tab:samples}
\end{table}

\section{Event and object selection}
\label{sec:sel}
 
This analysis uses 139~\ifb\ of data from 13~\TeV\ proton--proton collisions with 25~ns bunch spacing
collected by ATLAS from 2015 to 2018.
The data were subjected to quality requirements~\cite{DAPR-2018-01}, including the removal of events recorded when relevant detector components were not operating correctly.
Events for this search were selected  with the un-prescaled \met trigger with the lowest threshold~\cite{TRIG-2019-01}.
In 2015, a threshold of 70 \GeV\ was used; this was subsequently raised to cope with increasing effects
from pile-up as the LHC achieved higher luminosities, reaching 110~\GeV\ during the 2016--2018 data-taking period.
The \met in the trigger is based only on calorimetric measurements and does not
include any reconstructed muons, so the muons behave similarly to invisible particles in this trigger.
Events must have at least one reconstructed vertex
with at least two associated tracks with $\pT > 500$~\MeV.
The vertex with the highest sum of the squared transverse momenta of associated tracks is taken as the primary vertex.
 
Particle-flow (PFlow) jets are constructed using the \antikt algorithm~\cite{Cacciari:2008gp,Fastjet}
with a radius parameter of $R = 0.4$, using charged constituents associated with the primary
vertex and neutral PFlow constituents as inputs~\cite{PERF-2015-09}.
Jets are calibrated to the particle-level scale using a sequence of corrections,
including pile-up subtraction and energy and angular calibrations~\cite{JETM-2018-05}.
Events are required to have at least two jets within $|\eta| < 2.8$, the leading jet is required to
have $\pT > 250$~\GeV, and other jets are required to have $\pT > 30$~\GeV.
To suppress jets originating from pile-up collisions, requirements on the jet-vertex tagger~\cite{PERF-2014-03} discriminant are applied to jets with \pT below 60~\GeV.
Events are also required to have at least one jet within $\Delta \phi = 2.0$ of the $\vec{p}^\text{miss}_\text{T}$ direction.
The distance between the $\vec{p}^\text{miss}_\text{T}$ direction and the closest jet decreases with higher \Rinv\ fractions.
Jets are identified as \bjets if they pass the 77\%
efficiency working point of the DL1r algorithm~\cite{FTAG-2019-07}.
The \bjet-tagging performance has a strong dependence on the jet \pT.
The efficiency drops from 65\% for a \bjet \pT of around 500 \GeV to 10\% for a \pT of around 2 \TeV.
Estimated from MC simulation, the corresponding mistag rate of charm jets drops from 15\% to 2\% over the same \pT interval,
and that of light-flavour jets remains at the level of 1\%~\cite{FTAG-2019-07}.
Events with two or more \bjets are vetoed to reduce \ttbar background contributions.
A tighter veto cannot be applied because the signal topology contains some fraction of jets which originate from $b$-quarks.

Events with jets containing anomalous energy depositions due to coherent noise or
electronic noise bursts in the calorimeter~\cite{DAPR-2012-01, ATLAS-CONF-2015-029}
are removed, but this has a negligible effect on the signal efficiency.
Non-collision backgrounds, e.g.\ energy depositions in the calorimeters due to muons of
beam-induced or cosmic-ray origin, are suppressed by imposing an additional selection criterion on the leading jet:
the ratio of the jet charged-particle fraction to the maximum fraction of the jet energy collected
by a single calorimeter layer, $f_{\textrm{ch}}/ f_{\textrm{max}}$, is required to be larger than 0.1.
Events are rejected if they contain a selected jet
pointing in the direction where tile calorimeter modules were disabled,
or where other detector errors were present.
The non-collision backgrounds contribution is found to be negligible after these requirements.

Hadronically decaying $\tau$-lepton candidates are formed by using a Recurrent Neural Network algorithm
which combines information from the calorimeters and inner tracking detectors,
and they are required to have one or three associated tracks~\cite{ATL-PHYS-PUB-2019-033}.
Any events with such a $\tau$-lepton candidate with $\pT > 20$~\GeV\ and $|\eta| <$ 2.5 are rejected.
Electron candidates are reconstructed from energy clusters in the electromagnetic calorimeter
that are associated with charged-particle tracks reconstructed in the inner detector.
Electrons are required to fulfill \enquote{tight likelihood} identification criteria
as well as calorimeter- and track-based
isolation criteria~\cite{EGAM-2018-01}.
Muon candidates are reconstructed by combining inner-detector tracks with muon spectrometer tracks
or energy deposits in the calorimeters consistent with the passage of muons.
Muons are required to fulfil \enquote{medium} identification criteria as well as calorimeter- and track-based
isolation criteria~\cite{PERF-2015-10}.
Electrons and muons are required to satisfy $\pT > 7$~\GeV\ and be within the tracking
volume $|\eta| < 2.5$. Electrons are also required not to
be in the transition region $1.37 < |\eta| < 1.52$ between the barrel and endcap EM calorimeters.
 
An overlap removal procedure is applied to avoid ambiguities in reconstructing the objects specific to this final state.
Firstly, any electron which shares a track with a muon is rejected. Any jet whose angular distance $\Delta R$ from
an electron is less than 0.2 is removed, as is any which has fewer than three tracks and lies within $\Delta R = 0.4$ of
a muon. Finally, electrons and muons within $\Delta R = 0.4$ of the remaining jets are discarded.

The missing transverse momentum vector, $\vec{p}^\text{miss}_\text{T}$ is reconstructed as the negative vector sum of
transverse momenta of all selected
objects, as well as tracks compatible with the primary vertex but not matched to any of those objects;
this last contribution is called the \enquote{soft term}~\cite{PERF-2016-07}. The magnitude
of $\vec{p}^\text{miss}_\text{T}$ is referred to throughout this Letter as the missing transverse
momentum, denoted by \met.
Events with offline $\met > 250$~\GeV\ are selected in order to ensure that the trigger is fully efficient.
If muons are selected in the event, the $\vec{p}^\text{miss}_\text{T}$ is recalculated, considering muons to be invisible, to mimic an invisibly decaying
$W$ or $Z$ boson, in order to have a consistent definition of $\vec{p}^\text{miss}_\text{T}$ in the signal and control regions.
The lepton and jet momentum scales and resolutions, the lepton
reconstruction, identification, and isolation efficiencies, the \bjet identification efficiency, and the trigger efficiency in the simulation are corrected to match
those measured in data.

\section{Analysis strategy}
\label{sec:ana}
 
After the preselection of events with $\met > 250$~\GeV\ and at least two jets,
with at least one being within $\Delta \phi = 2.0$ of the $\vec{p}^\text{miss}_\text{T}$ direction,
the signal region (SR) is defined with $\met > 600$~\GeV\ and $\HT > 600$~\GeV, where
\HT is the scalar sum of the \pT of jets in the event. As mentioned earlier, events with at most
one \bjet are selected in SR.
The reason for this is twofold: signal events with higher mediator masses and \Rinv\ fractions
typically have high \met, and the background contribution from mismeasured multijet events
becomes subdominant.
Events with any electrons or muons passing the $\pT > 7$~\GeV\ requirement are discarded in the SR selection.
The dominant background contributions are then from processes with real $\vec{p}^\text{miss}_\text{T}$,
namely $W/Z$+jets, diboson and semileptonic top-quark processes.
The background from \ttbar and $W$+jets arises either because an electron or a muon is not observed
(because it has $\pT < 7$~\GeV, fails the identification requirement,
or is produced outside the central region of the detector)
or because a hadronically decaying $\tau$-lepton is misidentified as a jet.

In order to estimate the backgrounds resulting from these processes,
control regions (CRs) with the same \met and \HT requirements as the SR are defined using muons.
The 1L CR requires exactly one muon
and no \btagged jet. The 1L1B CR requires exactly one muon as before,
but also exactly one \btagged jet. Finally, the 2L CR requires
two oppositely charged muons with the pair's invariant mass lying between 66~\GeV\ and 116~\GeV,
and no \btagged jets.
The 1L CR is dominated by $W$+jets events, the 1L1B CR is
dominated by semileptonic \ttbar and single-top-quark processes as well as $W$+heavy-flavour jets processes, and the 2L CR
almost exclusively contains $Z$+jets events.
The CRs have negligible signal contamination.
 
The search then makes use of two other key observables, which are found to be largely uncorrelated:
 
\begin{enumerate}
 
\item the \pT balance between the jets closest ($j_1$) and farthest ($j_2$) in azimuth from the $\vec{p}^\text{miss}_\text{T}$ direction, denoted by \pTbal\ and defined using the two-dimensional \pT vectors:
\begin{equation*}
\pTbal =  \frac{|\vec{\pT}(j_{1})+\vec{\pT}(j_{2})|}{ |\vec{\pT}(j_{1})| + |\vec{\pT}(j_{2})|},
\end{equation*}
 
\item the azimuthal separation between $j_1$ and $j_2$ as defined above, denoted by \maxminphi.
 
\end{enumerate}

\begin{figure}
\centering
\subfloat[]{\includegraphics[width=0.45\textwidth]{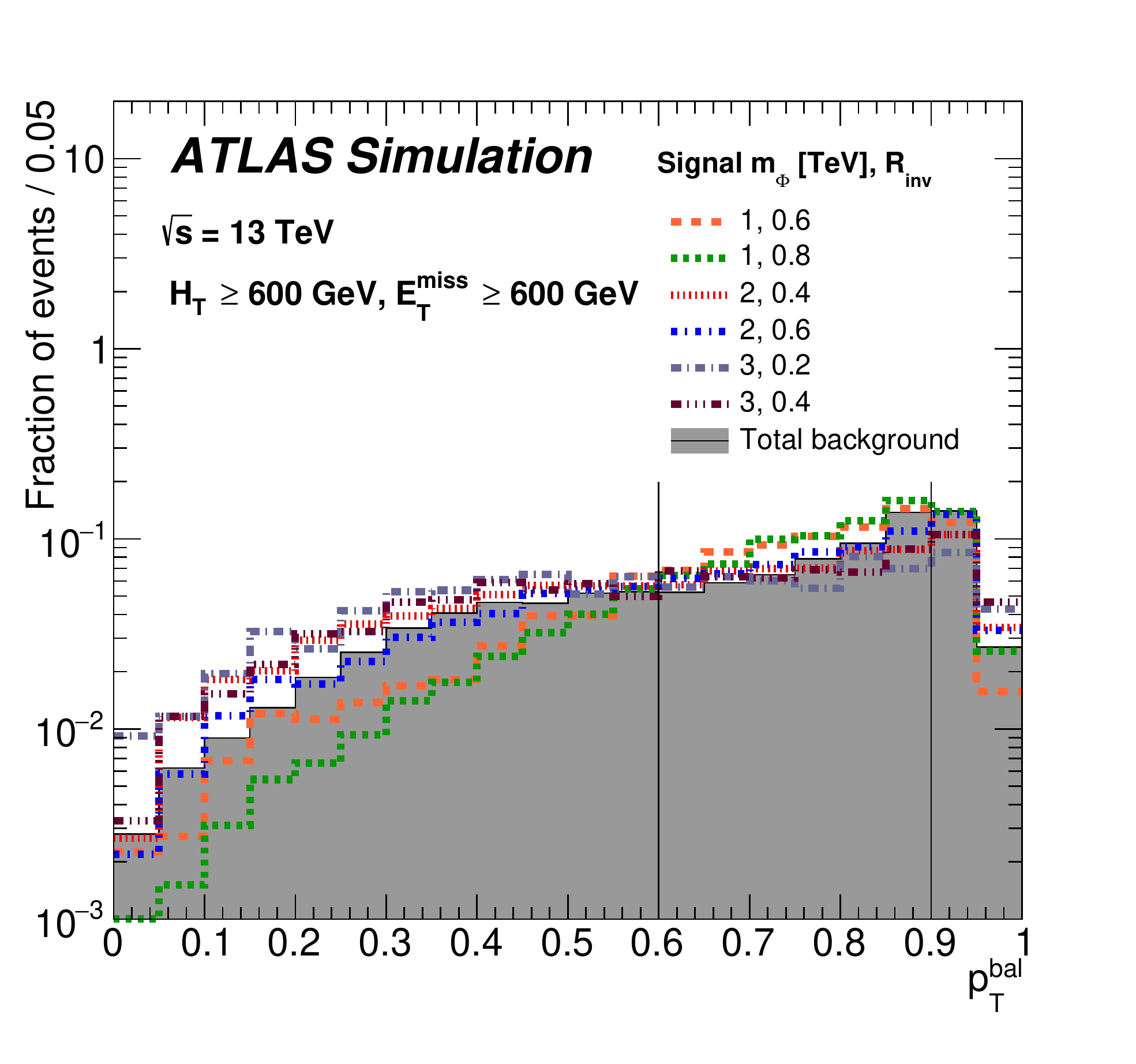}\label{}}
\hskip2ex
\subfloat[]{\includegraphics[width=0.45\textwidth]{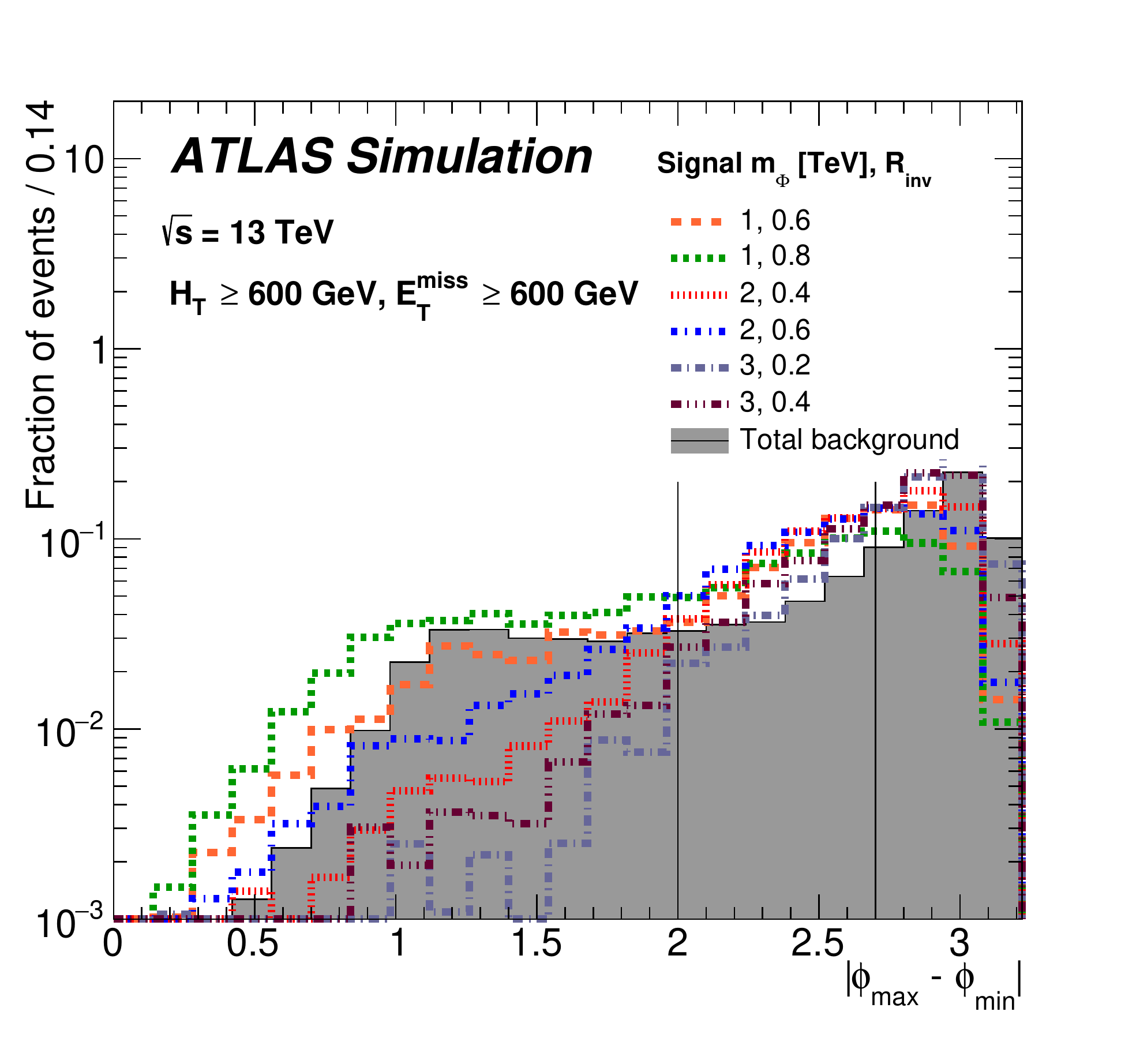}\label{}}
\caption{Shape comparisons of the (a) \pTbal\ and (b) \maxminphi\ distributions
of the total background before the fit and six signal predictions for
representative mediator masses and invisible fractions in SR.
The solid vertical lines show how these
distributions are subsequently divided to form the nine-bin grid.}
\label{fig:inclkin}
\end{figure}
 
Figure~\ref{fig:inclkin} compares the shapes of the signal and total background  \pTbal\ and \maxminphi\ distributions in the SR.
There are distinct distribution shape differences between the total background and the different signal benchmark points,
and these are utilised in designing the fit strategy.
The \maxminphi\ distribution is divided into three bins covering $<2$, 2--2.7, and $>2.7$,
and the \pTbal\ distribution is divided into three bins covering $<0.6$, 0.6--0.9, and $>0.9$.

Although the simulated signals mostly populate the higher part of the \pTbal\ and \maxminphi\ ranges,
these regions also have larger background contributions and both have less signal sensitivity than the lower part of the \maxminphi\ distribution.
The nine bins are defined identically in the SR and CRs.
The yields in these bins of \pTbal\ and \maxminphi\ are treated as the observables used in the fit described in Section~\ref{sec:fit}.

For the two sensitive observables chosen above, mostly good agreement between data and MC predictions is observed in the various CRs.
However, since no independent CR for multijet
processes can be established, an additional low-\met\ multijet reweighting region (MJRR)
for multijet processes is defined by requiring \met to be between
250~\GeV\ and 300~\GeV\ with the same $\HT > 600$~\GeV\ requirement as in the SR and CRs.
In this MJRR, the \pTbal\ bins in the range $2.7 < \maxminphi < 3.2$ are found to be
multijet-rich. In these bins, the contribution from multijet processes is estimated
by subtracting the other predicted SM process yields from the data event
yield after applying the scale factors obtained for these backgrounds in Section~\ref{sec:fit}.
The differences observed between this predicted multijet yield and MC predictions
for multijet processes determine the reweighting factors which are applied to
the other \maxminphi\ bins in the same \pTbal\ range.
This reweighting factors range from 2 to 2.4 for different \pTbal\ ranges.
Tests showed that using different bin-ranges causes only small changes in those factors,
and their values also remained compatible when derived from ranges of lower or higher \MET values.
A fit is initially performed using both SRs and CRs to extract the multijet reweighting factors. Once the reweighting factors are derived, the final
CR-SR simultaneous fit described in Section~\ref{sec:fit} is performed after accounting for the obtained factors.

\section{Statistical analysis and background estimation}
\label{sec:fit}
 
In order to estimate the signal strength, as well as to better determine the individual
yield $N_{i}^{\textrm{bg}}$ of the $i$-th background having a probability distribution function $P_{i}$,
a simultaneous binned maximum-likelihood function fit is performed on the product of
all $P_{i}$ and the nine bin yields, using MC templates, by including both the SR and the corresponding CRs (1L, 1L1B, and 2L). This provides a way to search for the signal while simultaneously improving the background prediction in the SR.
The normalisation factors $k_i^{\textrm{SF}}$ for the individual backgrounds are allowed to float and are determined from the fit:
\begin{equation*}
\mathcal{L}(\mu, \theta) = \prod_{j \in 36~\textrm{bins}} \textrm{Poisson}(N_j^{\textrm{obs}} | \mu N_j^{\textrm{sig}}(\theta) +
\sum_{i \in \textrm{bg}} k_i^{\textrm{SF}} \times N_{i,j}^{\textrm{bg}}(\theta) ) \times f^{\textrm{constr}}(\theta)
\label{eqn:likhood}
\end{equation*}
where $N_j^{\textrm{obs}}$ is the observed total yield, $N_j^{\textrm{sig}}(\theta)$
is the total predicted signal yield, and $N_{i,j}^{\textrm{bg}}(\theta)$ is the predicted background yield for the $i$-th background.
The signal strength $\mu$ is given by the ratio of the measured to predicted signal cross-section.
All the systematic uncertainties discussed in Section~\ref{sec:syst} are propagated into the simultaneous fit
as different nuisance parameters (NPs), denoted by $\theta$.
The term $f^{\textrm{constr}}(\theta)$ represents the product of the Gaussian constraints applied to each of the nuisance parameters,
and is defined as
\begin{equation*}
f^{\textrm{constr}}(\theta) = \prod_{k = 1}^M G_k(\theta)
\end{equation*}
where $G_k$ is the standard normal distribution, and $M$ is the total number of systematic uncertainty sources.
The MC templates are allowed to vary within their shape uncertainty, and the NPs representing the systematic uncertainties are correlated across bins and signal and control regions unless stated otherwise.
The fit finds the set of values of the unknown parameters $\mu$ and $\theta$ that maximises
${\mathcal{L}}$.
The profile-likelihood ratio is used as the test statistic, and upper limits on the
contribution of events from new physics are computed by using the modified frequentist approach $CL_s$
based on asymptotic formulas at 95\% confidence level~\cite{Cowan:2010js}.
It should be noted that $\mu$ is allowed to take negative values in the fit, and $CL_s$ deals with this
value by considering a different likelihood ratio than that which would be utilised for positive-definite $\mu$ values.
The different likelihood ratio then corrects for any over-constraining of the limits.
The fit also finds the
uncertainty in $\mu$ by taking into account the correlations between all sources of uncertainty.
The post-fit scale factors for different background processes are listed in Table~\ref{tab:sfcomp2}.
The normalisation scale factors have a correlation of 20\% between the top-quark and $W$+jets processes,
14\% between $W$+jets and $Z$+jets processes, and 5\% between top-quark and $Z$+jets processes.

\begin{table}[ht]
\centering
\caption{Scale factors for each background process
obtained from the simultaneous fit using the SR, 1L CR, 1L1B CR and 2L CR.
\enquote{Top processes} denotes merged contributions from \ttbar and  single top-quark processes.
The quoted uncertainties include statistical and systematic uncertainties.
As can be seen, the multijet process normalisation factor is close to unity after the
multijet reweighting procedure. The $W$+jets and $Z$+jets process scale factors are primarily driven
by the dedicated control regions and they are close to unity as well.}
\begin{tabular}{lc}
\hline
Process & $k^{\textrm{SF}}$ \\
\hline
$Z$+jets & 1.18 $\pm$ 0.05 \\
$W$+jets & 1.09 $\pm$ 0.04 \\
Top processes & 0.64 $\pm$ 0.04\\
Multijet & 1.10 $\pm$ 0.04\\
\hline
\end{tabular}
\label{tab:sfcomp2}
\end{table}

\FloatBarrier

\section{Systematic uncertainties}
\label{sec:syst}
 
Systematic uncertainties in the signal and background yields and shapes
result
from experimental uncertainties and theoretical modelling effects.
Except the luminosity uncertainty, all the other uncertainties are shape uncertainties, but eventually they are translated to 9-bin yields.
The experimental ones are due to the jet energy scale (JES) and resolution (JER)~\cite{PERF-2016-04},
computation of the \met soft term~\cite{PERF-2016-07},
flavour-tagging performance~\cite{FTAG-2018-01},
rescaling of simulation to match the pile-up profile in data,
and an uncertainty in the luminosity estimation.
The uncertainty in the combined 2015--2018 integrated luminosity is 1.7\%~\cite{ATLAS-CONF-2019-021},
obtained using the LUCID-2 detector~\cite{Avoni_2018} for the primary luminosity measurements.
The luminosity uncertainty is applied to background processes that are normalised to
theoretical predictions and to the signal cross-section parameters in the fit.
Uncertainties in the reconstruction, identification, isolation~\cite{EGAM-2018-01,MUON-2018-03, ATLAS-CONF-2017-029}
and trigger efficiencies~\cite{TRIG-2018-01} of muons, electrons and
$\tau$-leptons and in their energy scale and resolution are also considered.

Theoretical uncertainties common to the MC samples are due to renormalisation
and factorisation scale choices, and also the choice of PDF set and value for the strong coupling constant \alphas.
These uncertainties were assessed from the effect of varying the scales, the PDFs, and the value of \alphas.
Additionally, initial- and final-state radiation scale uncertainties were assessed for
\ttbar and single-top processes, and were taken to be uncorrelated.
For \ttbar and $tW$ processes,
alternative predictions were obtained by using \AMCatNLO
as the matrix element generator, also interfaced to \Pythia[8.230],
or by using the same matrix element generator \POWHEGBOX[v2],
but interfaced instead to \Herwig[7]~\cite{Bahr:2008pv,Bellm:2015jjp} for the modelling of the parton shower, hadronisation, and underlying event.
The differences are treated as systematic uncertainties in the modelling of these processes.
Because the fit is performed using the same \HT requirement in the CRs and the SR,
and the observables most sensitive to signal are largely insensitive to the \pT of the top quark,
no additional uncertainty related to the modelling of the top quark's \pT is considered.
The difference between using the diagram-subtraction and diagram-removal schemes~\cite{Frixione:2008yi} to remove
the overlap between the $tW$ and \ttbar processes is also considered.
The $W$+jets samples were split into heavy- and light-flavour subprocesses,
and since the MC prediction was found to underestimate the former
by about a factor of 1.3~\cite{HIGG-2016-29}, an additional 30\% normalisation uncertainty is assigned to the heavy-flavour samples.
Scale variations are treated as uncorrelated between the $W$+jets heavy- and light-flavour subprocesses.
 
An additional uncertainty from the reweighting procedure is also considered for the multijet modelling,
and is conservatively assessed to be 100\% for each of the bins, except those used to obtain the reweighting factors
as described at the end of Section~\ref{sec:ana}.
This uncertainty is treated as fully uncorrelated between the bins,
and was checked in an
intermediate-\met validation region of 300--600~\GeV\ with the same $\HT > 600$~\GeV\ requirement.
The agreement of the data with the sum of the predicted background indicates
no additional mismodelling of the multijet background beyond the quoted uncertainty.

\section{Results}
\label{sec:res}
 
The post-fit yields for all nine bins are shown in Figure~\ref{fig:9bin} of the three CRs and the SR separately,
and listed in Table~\ref{tab:binyield}.
Overall, good agreement between data and SM predictions is observed for all the bins.
The post-fit distributions of the \HT, \met, \maxminphi\ and \pTbal\ observables
are shown in Figure~\ref{fig:kin} for the SR. Again, excellent agreement of data
with SM background predictions is seen for all the observables, which also indicates
the absence of any effect beyond the SM.
The largest post-fit effects on the shapes of discriminating observables are signal modelling uncertainties of up to 8\%, $Z$+jets modelling uncertainties of up to 7\%, and top-quark process modelling uncertainties of up to 4\%. The rest of the contributions are less than 2\%.
The total data yield can be translated into a model-independent cross-section
limit in this SR.

\begin{figure}
\centering
\subfloat[]{\includegraphics[width=0.48\textwidth]{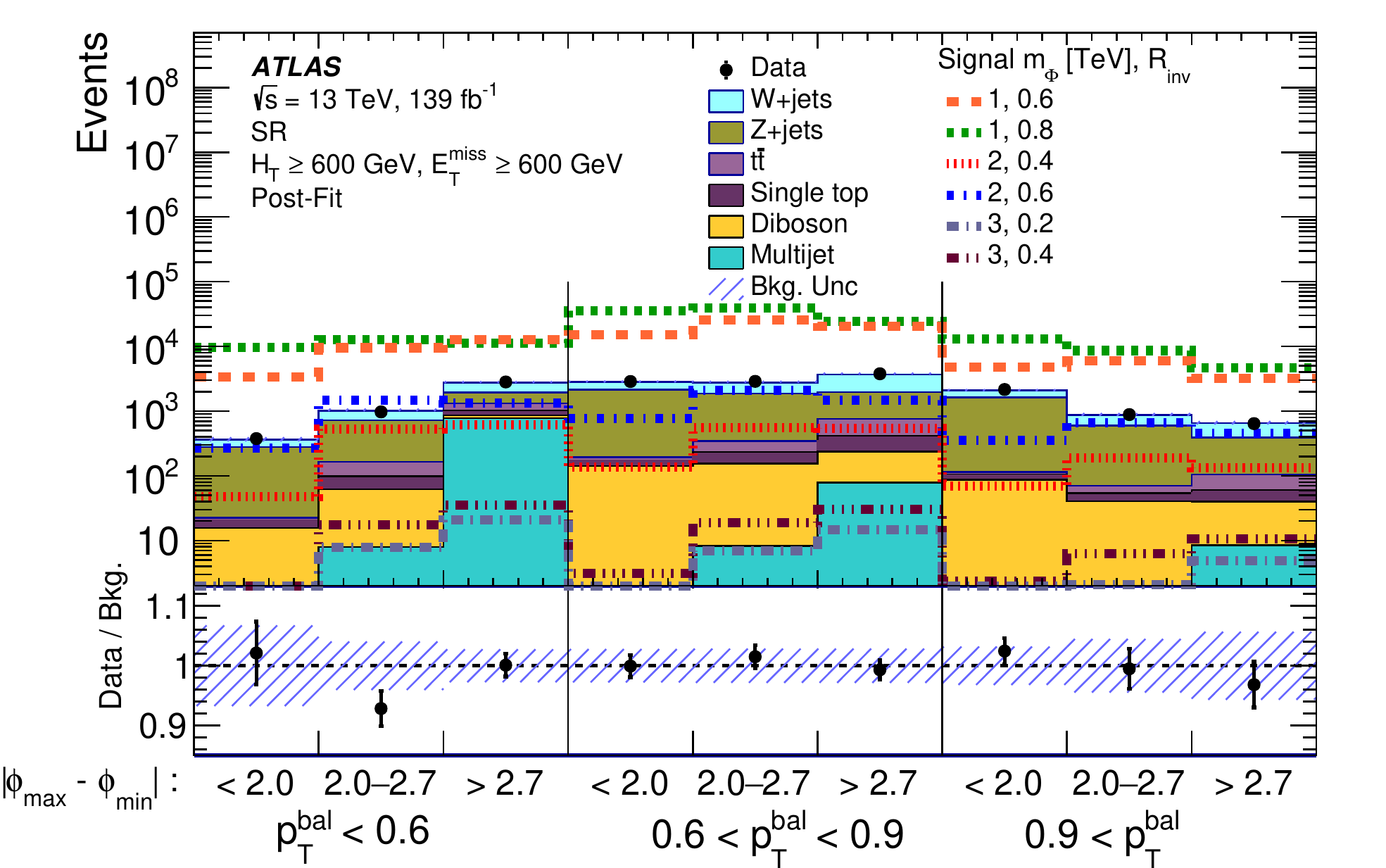}\label{}}
\hskip2ex
\subfloat[]{\includegraphics[width=0.48\textwidth]{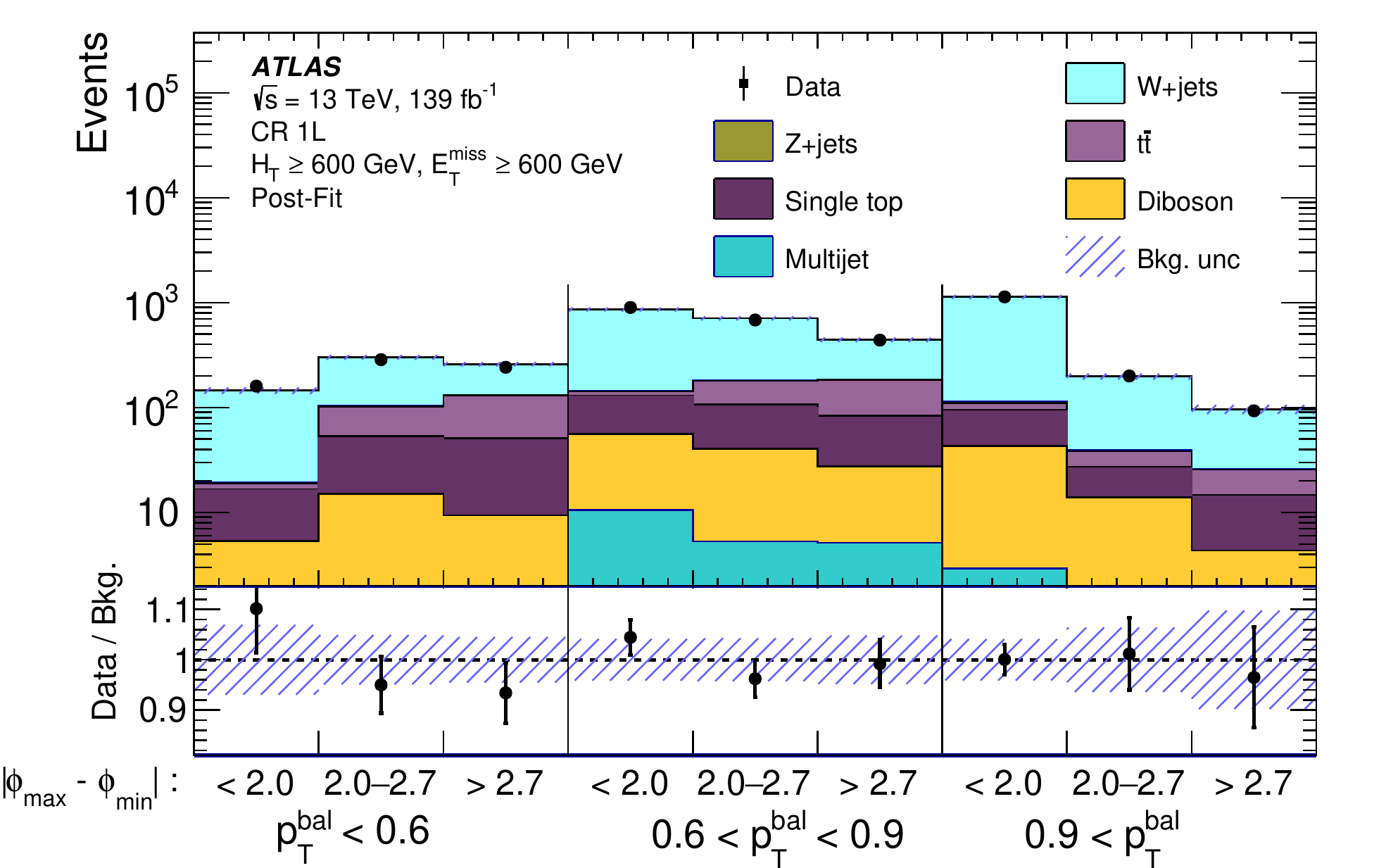}\label{}} \\
\subfloat[]{\includegraphics[width=0.48\textwidth]{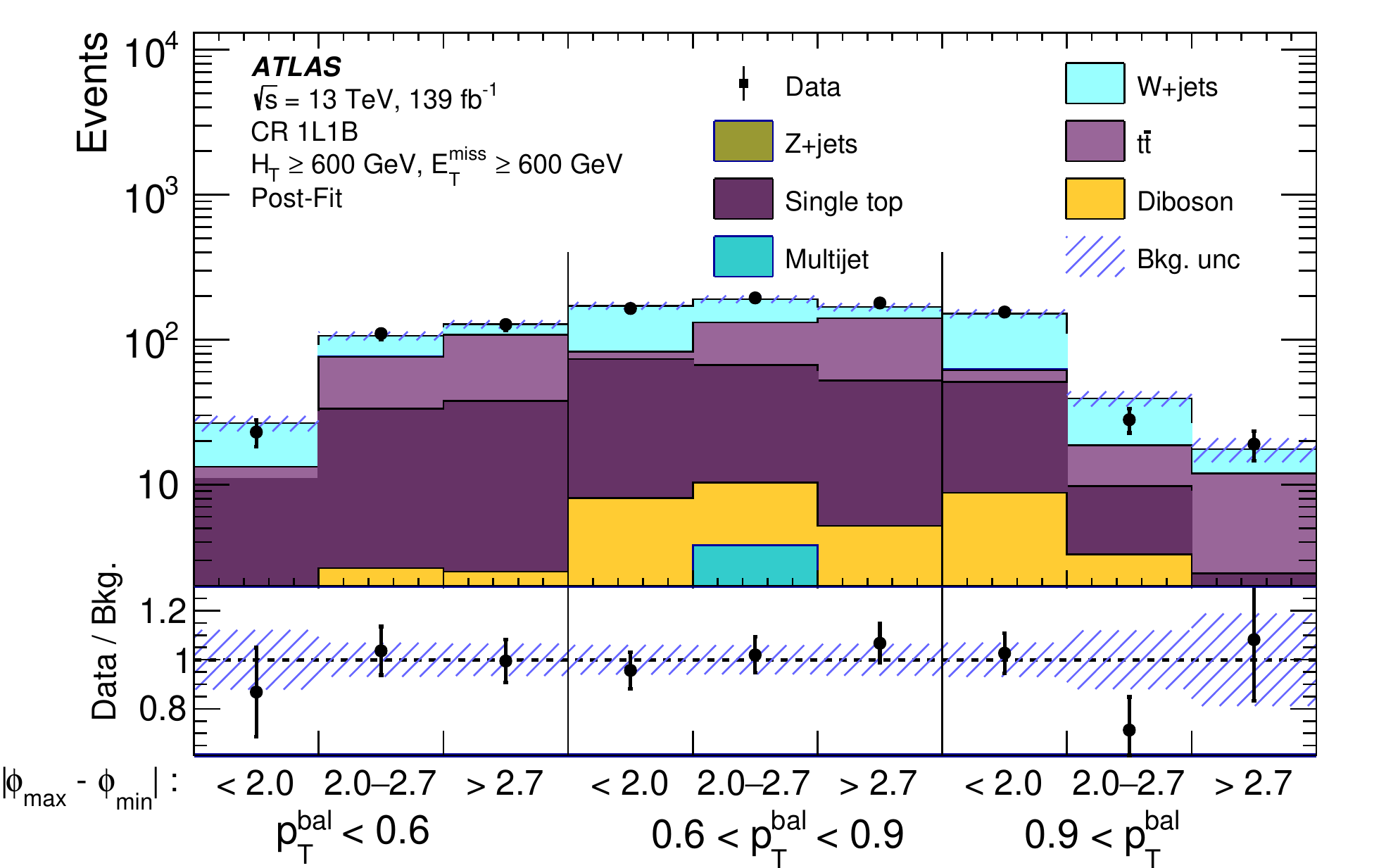}\label{}}
\hskip2ex
\subfloat[]{\includegraphics[width=0.48\textwidth]{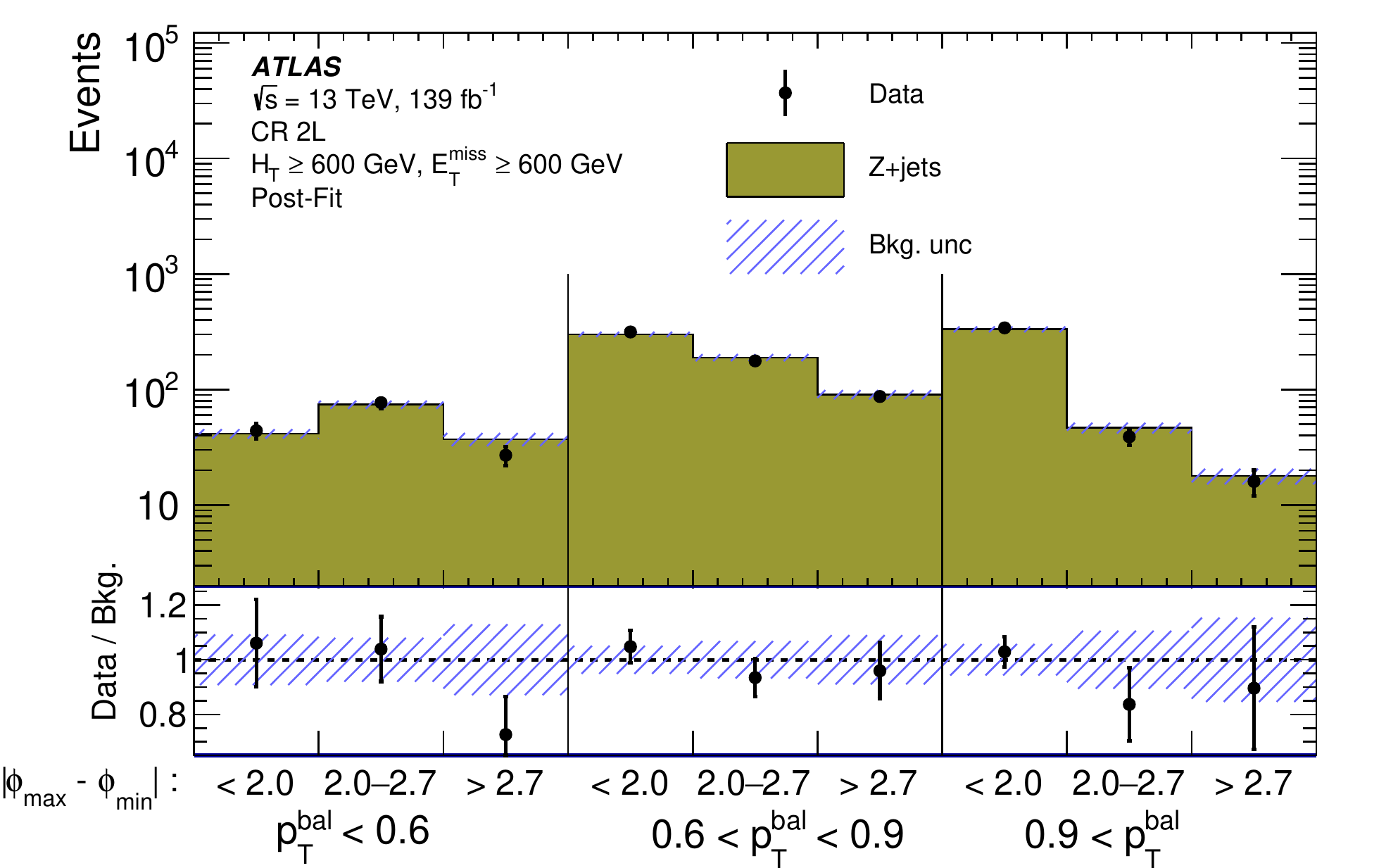}\label{}} \\
\caption{The post-fit yields in the nine bins of the $(\pTbal, \maxminphi)$ grid
are shown for the (a) SR, (b) 1L CR, (c) 1L1B CR, and (d) 2L CR.
Data are compared with background predictions,
and six signal predictions covering a representative
mediator mass and invisible fraction range are overlaid in the SR.
Figure (a) shows a background-only fit in the SR. 
The uncertainties include all systematic and statistical components. }
\label{fig:9bin}
\end{figure}

\begin{table}[ht]
\centering
\caption{Post-fit yields from the background-only fit,
including pre-fit contributions of different signal benchmark points.
Dashes refer to components that are negligible or not applicable.
The total uncertainties include statistical and systematic uncertainties.}
\begin{tabular}{lr@{$~\pm~$}lr@{$~\pm~$}lr@{$~\pm~$}lr@{$~\pm~$}lr@{$~\pm~$}lr@{$~\pm~$}lr@{$~\pm~$}lr@{$~\pm~$}l} 
\toprule
Process	        &  \multicolumn{2}{c}{SR}	& \multicolumn{2}{c}{CR 1L}	& \multicolumn{2}{c}{CR 1L1B}	& \multicolumn{2}{c}{CR 2L}\\
\midrule
$Z$+jets		&  8\,490 &  260 & 11.6 &  1.4 & 2.2 &  0.6 & 1\,120 &  40  \\
$W$+jets	    &  5\,820 &  300 & 3\,190 &  170 & 351 &  41 & \multicolumn{2}{c}{-}  \\
\ttbar		&  920 &  70 & 350 & 29 & 304 &  24 & \multicolumn{2}{c}{-} \\
Single top	&  533 &  47 & 358 & 29 & 290 &  25 & \multicolumn{2}{c}{-}\\
Multijet	    &  850 & 100 & 28 & 11 & 7.7 & 3.1 & \multicolumn{2}{c}{-}  \\
Diboson		&  757 & 10 & 187 & 9 & 34.5 & 2.8 & \multicolumn{2}{c}{-} \\
\midrule
Total bkg.	&  17\,370 &  280 & 4\,120 &  100 & 990 &  35 & 1\,120 &  40\\
\midrule
Data	        &  \multicolumn{2}{c}{17\,388} & \multicolumn{2}{c}{4\,136} & \multicolumn{2}{c}{999} & \multicolumn{2}{c}{1\,124} \\
\midrule
Signal: &  \multicolumn{4}{c}{}  \\
\Mphi~=~1~\TeV , \Rinv~=~0.6	    &  101\,000 & 23\,000 & \multicolumn{2}{c}{-} & \multicolumn{2}{c}{-} & \multicolumn{2}{c}{-} \\
\Mphi~=~1~\TeV , \Rinv~=~0.8	    &  160\,000 & 40\,000 & \multicolumn{2}{c}{-} & \multicolumn{2}{c}{-} & \multicolumn{2}{c}{-} \\
\Mphi~=~2~\TeV , \Rinv~=~0.4	    &  2\,800 & 600 & \multicolumn{2}{c}{-} & \multicolumn{2}{c}{-} & \multicolumn{2}{c}{-} \\
\Mphi~=~2~\TeV , \Rinv~=~0.6	    &  8\,900 & 2\,000 & \multicolumn{2}{c}{-} & \multicolumn{2}{c}{-} & \multicolumn{2}{c}{-} \\
\Mphi~=~3~\TeV , \Rinv~=~0.2	    &  59 & 13 & \multicolumn{2}{c}{-} & \multicolumn{2}{c}{-} & \multicolumn{2}{c}{-} \\
\Mphi~=~3~\TeV , \Rinv~=~0.4	    &  126 & 29 & \multicolumn{2}{c}{-} & \multicolumn{2}{c}{-} & \multicolumn{2}{c}{-} \\
\bottomrule
\end{tabular}
\label{tab:binyield}
\end{table}

Upper limits on the contribution of events from new physics are computed by using the modified
frequentist CL$_\text{s}$ approach based on asymptotic formulas at 95\% confidence level (CL)~\cite{Read:2002hq,Cowan:2010js}, and are
given in terms of the signal strength, $\mu$, defined previously.
For upper limits with values of $\mu \leq 1$, the nominal cross-section is excluded, while for
upper limits with $\mu > 1$, no such conclusion can be obtained.
The limits on the signal model are presented in two different ways.
The 95\% CL exclusion limit plots for limits on the cross-section as a function
of mediator mass and \Rinv\ values are shown in Figure~\ref{fig:contour}.
The observed limits increase from 2.4~\TeV\ for \Rinv\ of 0.2 to 2.7~\TeV\ for \Rinv\ of 0.8.
The observed exclusions tend to be slightly stronger than the expected ones due to a slight deficit
in data in individual SR bins and the fit's preference of a negative signal yield to improve the data agreement.
The exclusion confidence levels decrease at higher mediator masses as the distributions of the discriminating observables,
e.g.\ \maxminphi, become more background-like in shape and with lower yields.
The signal acceptance in the SR decreases for lower mediator masses, opposing the effect of rapidly increasing signal event rates.
The systematic uncertainties weaken the limits by about 25\%.

Additionally, the nominal signal cross-sections for each signal mass point, obtained with $\lambda=1$,
can be scaled by $\lambda^4$ for mediator masses larger than 2.5~\TeV. For each mediator mass point,
the limit on the cross-section is  obtained, and the corresponding $\lambda$ is calculated.
This $\lambda$ value corresponding to the cross-section upper limit is presented for the SR in Figure~\ref{fig:gridlimit}.
It can be seen that for lower mass points, the nominal cross-sections are excluded,
whereas for higher mass points only higher values of cross-sections can be excluded.
The advantage of this representation is that it sets stringent limits on the signature in general for a wide range of $\lambda$ values, and can help in recasting
this analysis for future model predictions.

\begin{figure}
\centering
\subfloat[]{\includegraphics[width=0.48\textwidth]{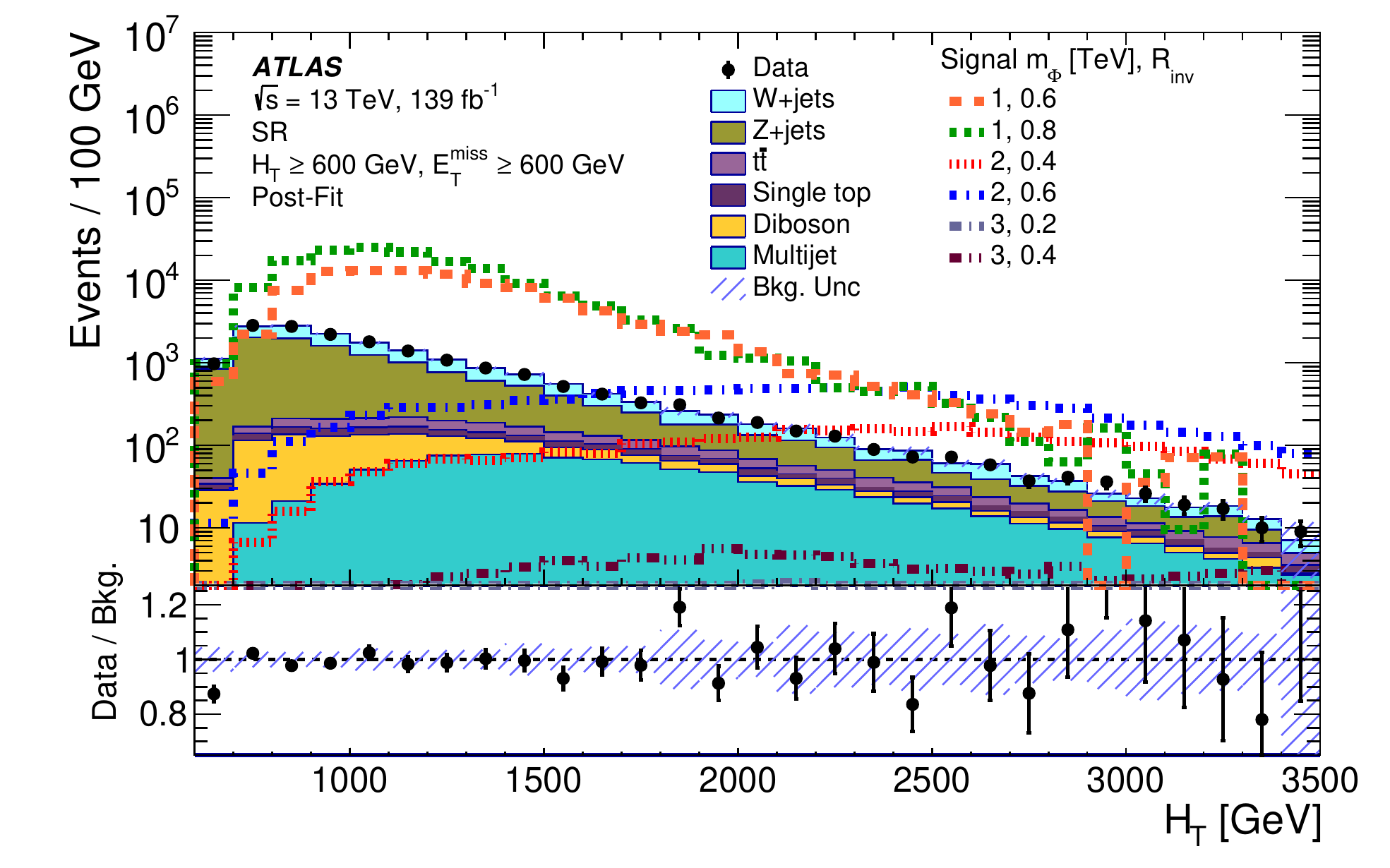}\label{}}
\hskip2ex
\subfloat[]{\includegraphics[width=0.48\textwidth]{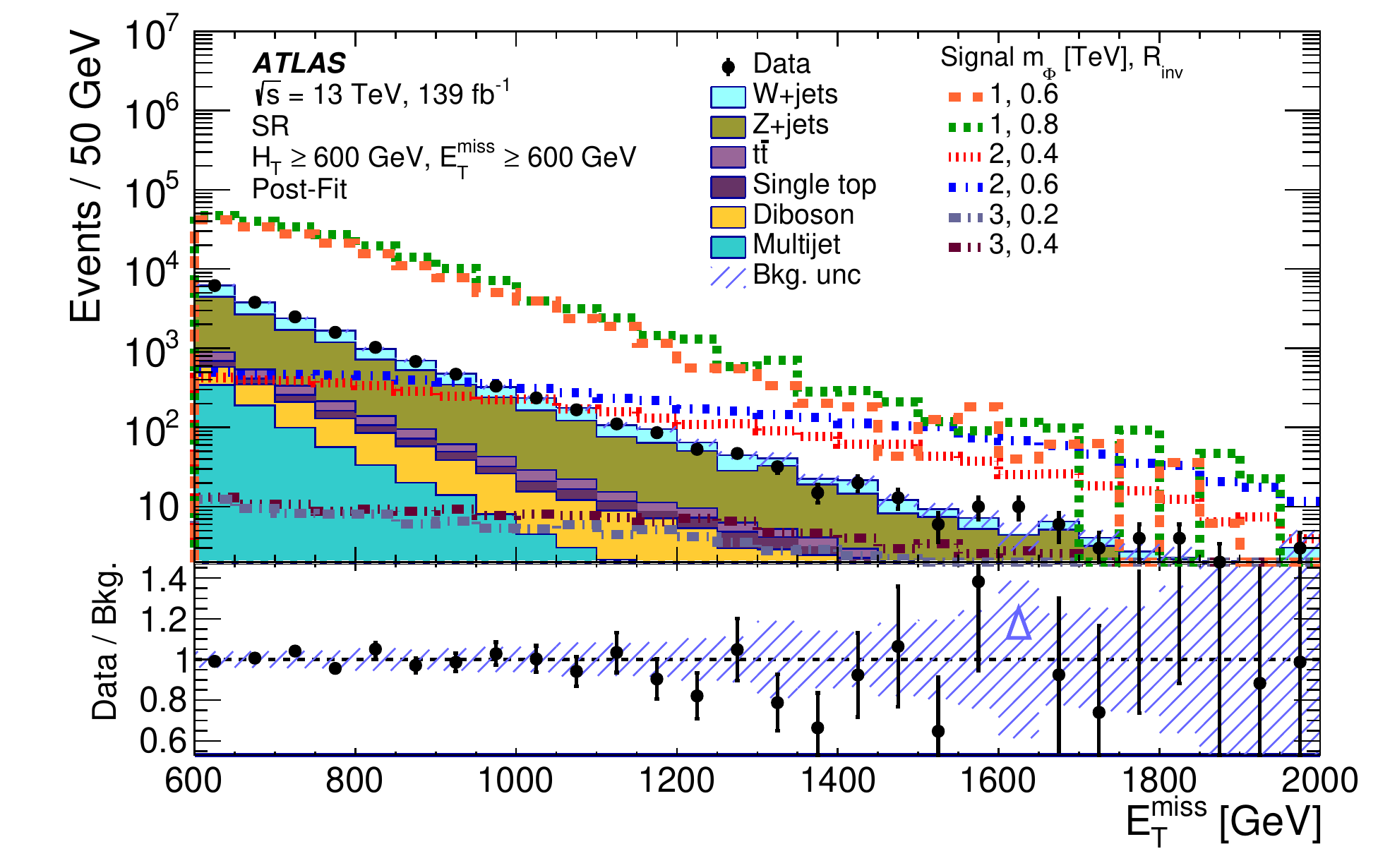}\label{}} \\
\subfloat[]{\includegraphics[width=0.48\textwidth]{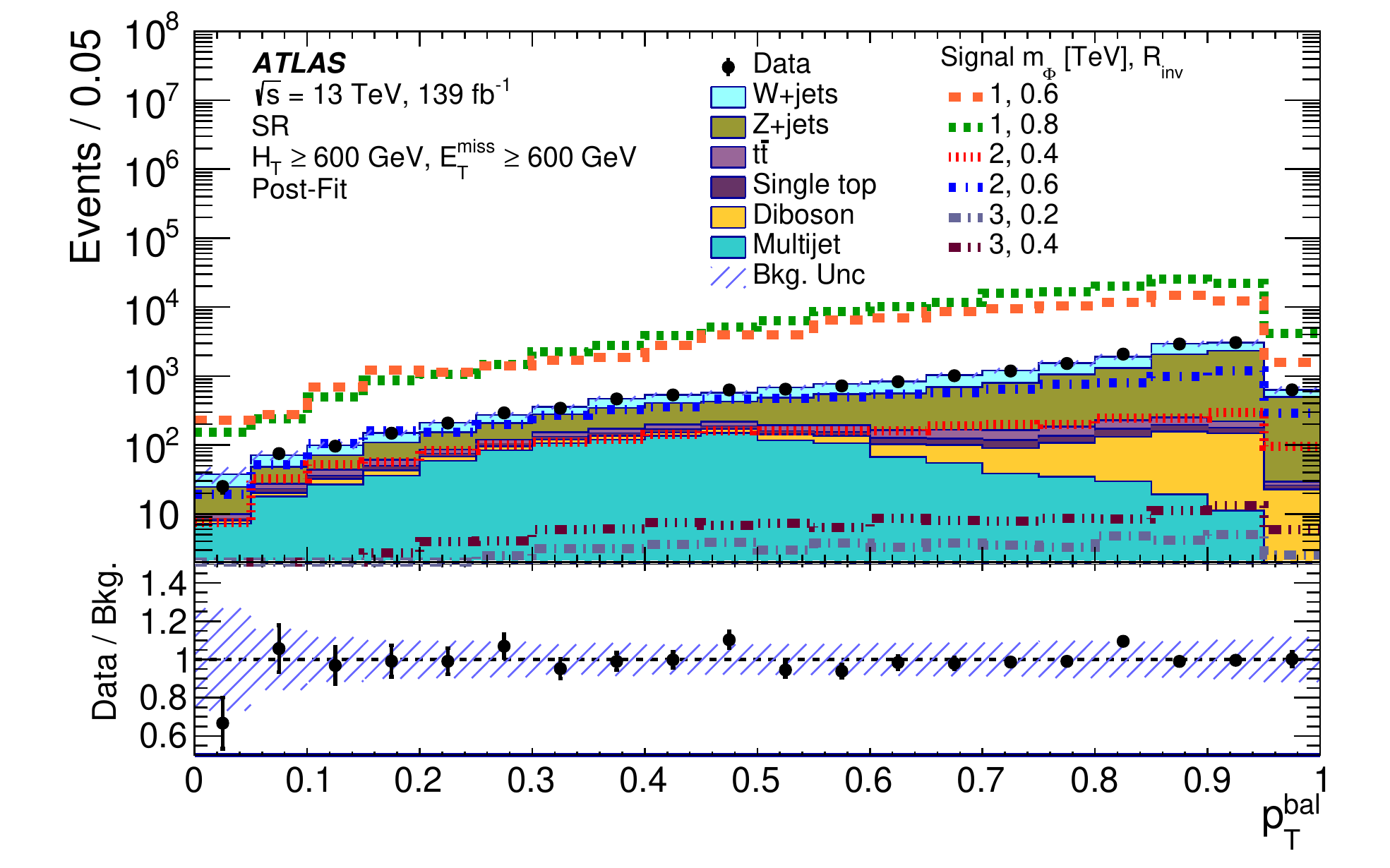}\label{}}
\hskip2ex
\subfloat[]{\includegraphics[width=0.48\textwidth]{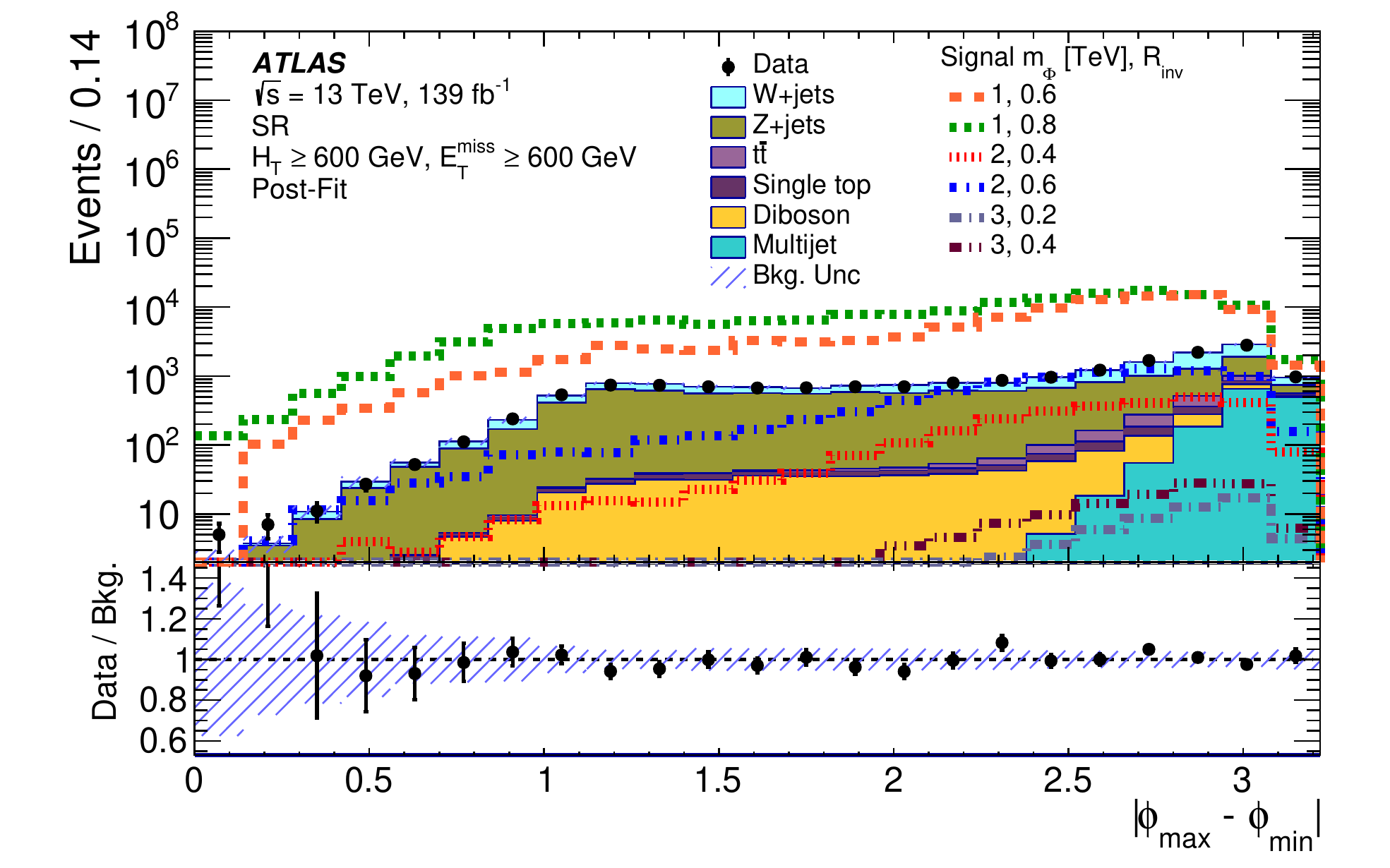}\label{}} \\
\caption{The post-fit distributions of (a) \HT, (b) \met, (c) \pTbal, and (d) \maxminphi\ are shown for the SR. Data are compared with background predictions, and six signal predictions covering a representative mediator mass and invisible fraction range are overlaid. The uncertainties include all systematic and statistical components. The last bin in (a) and (b) contains the overflow.}
\label{fig:kin}
\end{figure}

\begin{figure}[h]
\centering
\includegraphics[width=0.66\textwidth]{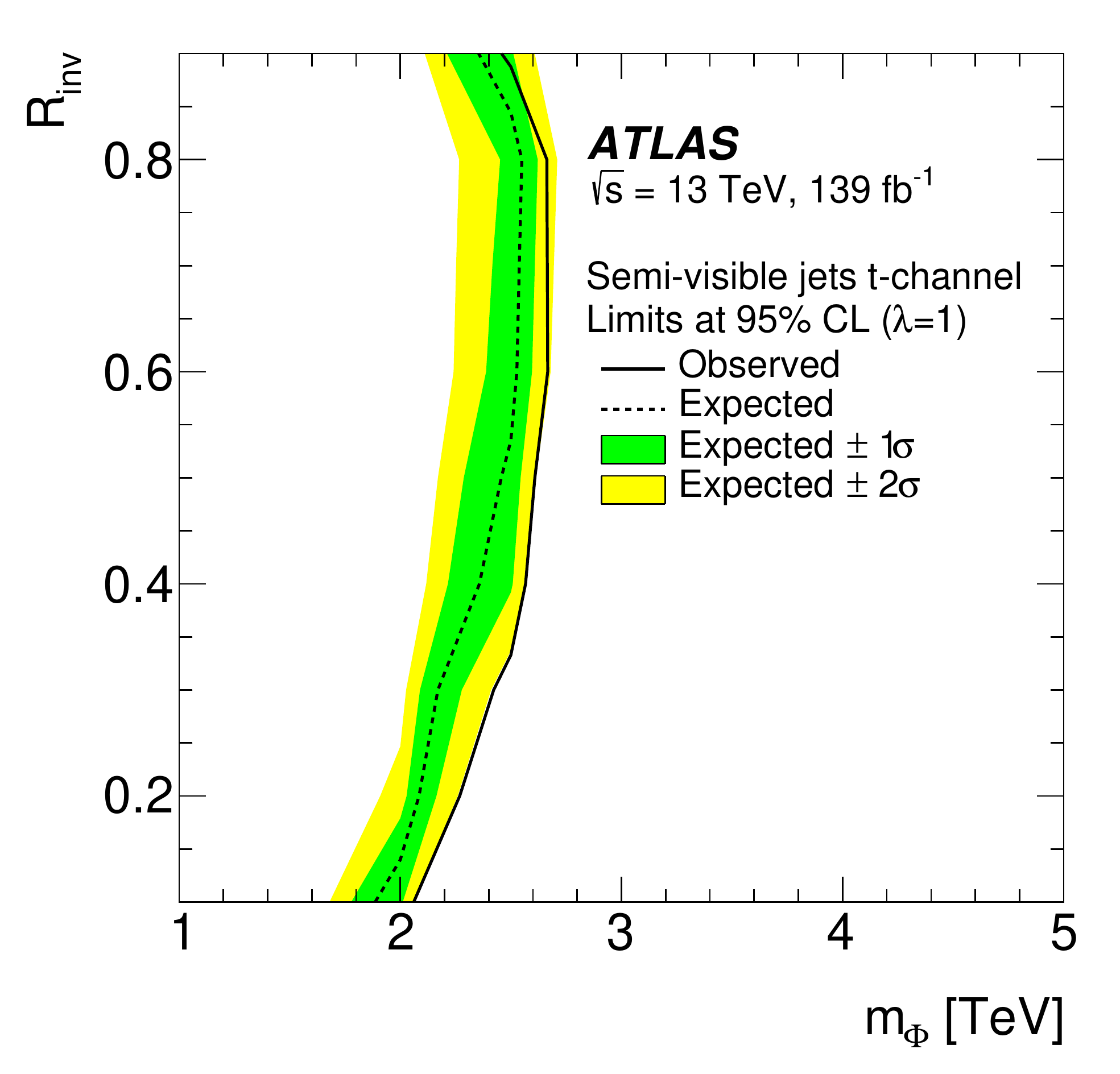}
\caption{
The expected and observed exclusion contours at 95\% CL for semi-visible jets signal as a function of mediator mass
\Mphi\ on the horizontal axis and \Rinv\ on the vertical axis. The solid black line is observed limit as a function of the mediator mass and \Rinv\. The green and yellow shaded bands correspond to the expected one and two standard deviation uncertainties, respectively, with the expected central value shown by the black dashed line.}
\label{fig:contour}
\end{figure}

\begin{figure}[ht]
\centering
\includegraphics[width=0.66\textwidth]{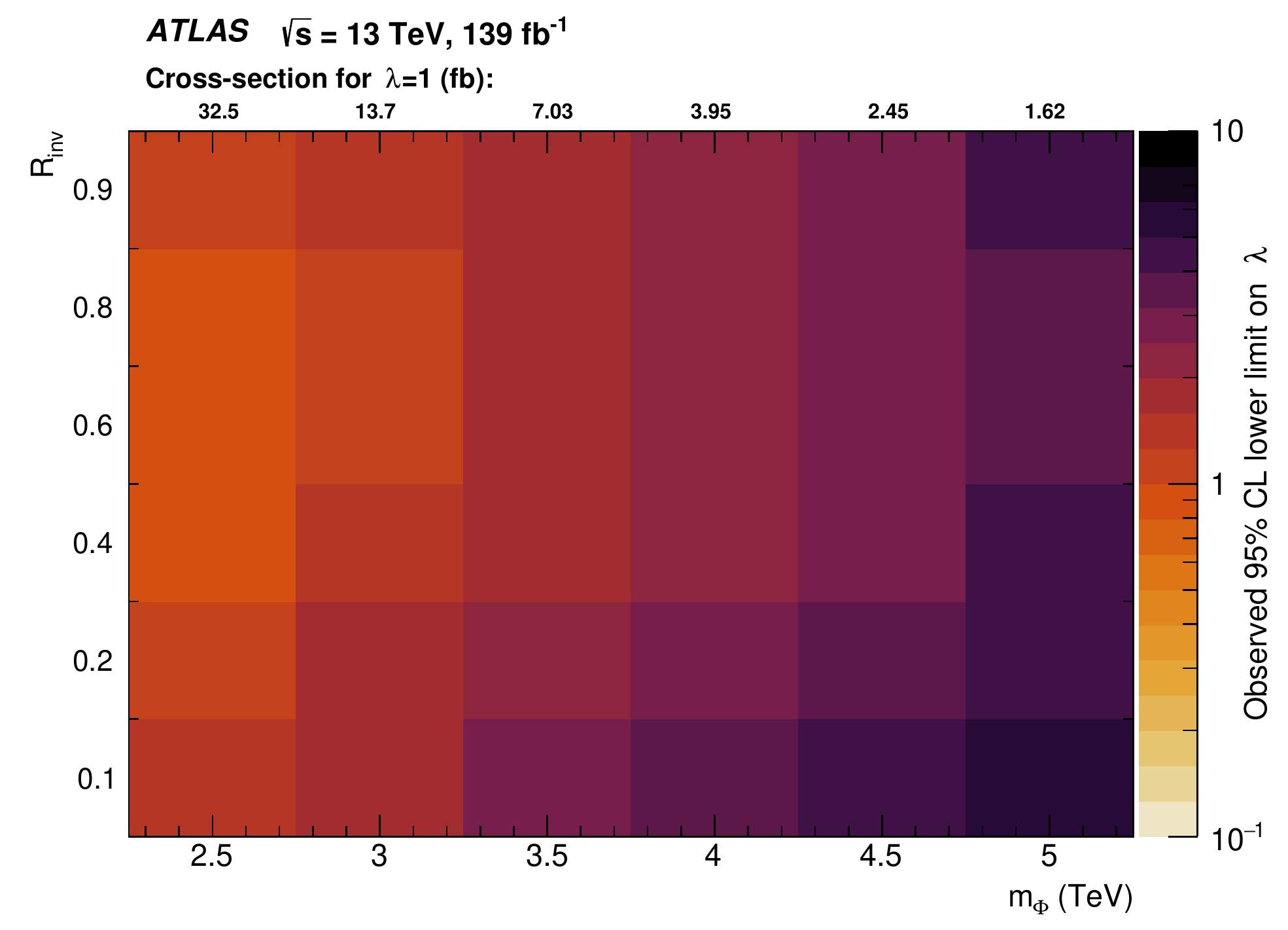}
\caption{The grid shows the observed 95\% CL lower limit on $\lambda$
with \Mphi\ on the horizontal axis, \Rinv\ on the vertical axis.
Shown above each \Mphi\ column is the predicted cross-section in fb for that
specific mass value as a reference.
As the nominal signal cross-sections for each signal mass point, obtained with $\lambda=1$,
can only be scaled by $\lambda^4$ for mediator masses larger than 2.5~\TeV, so only
mediator masses starting from 2.5~\TeV are shown.
}
\label{fig:gridlimit}
\end{figure}

\FloatBarrier

\section{Conclusions}
\label{sec:conc}

A first search is performed for semi-visible jets in the $t$-channel
production mode using 139~fb$^{-1}$ of data collected from $\sqrt{s} = 13$~\TeV\ proton-proton collisions
by the ATLAS experiment at the LHC.
Semi-visible jets arise in the strongly interacting dark sector,
where dark quarks are produced and their hadronisation generates jets
where dark hadrons
are interspersed with SM hadrons in the jet, leading to a novel collider signature.
The search considers mediator masses in the range 1--5~\TeV\ and
a stable-dark-hadron fraction, \Rinv, of  0.1--0.9.
The observed yields are in agreement with the SM background expectations
in the signal region defined by $\met > 600$~\GeV\
and $\HT > 600$~\GeV.
A total of 17388 data events are observed in the SR,
which can be translated into a model-independent cross-section upper limit relevant to the acceptance of this specific signal region selection.
The 95\% confidence-level upper limits on the mediator mass vary from 2.4~\TeV\ to 2.7~\TeV, depending
on the value of \Rinv. They are translated into upper limits on the coupling strength
between the mediator, a Standard Model quark and a dark quark.

\section*{Acknowledgements}

% The next lines are included from the .//acknowledgements/Acknowledgements.tex input file

We thank CERN for the very successful operation of the LHC, as well as the
support staff from our institutions without whom ATLAS could not be
operated efficiently.
 
We acknowledge the support of
ANPCyT, Argentina;
YerPhI, Armenia;
ARC, Australia;
BMWFW and FWF, Austria;
ANAS, Azerbaijan;
CNPq and FAPESP, Brazil;
NSERC, NRC and CFI, Canada;
CERN;
ANID, Chile;
CAS, MOST and NSFC, China;
Minciencias, Colombia;
MEYS CR, Czech Republic;
DNRF and DNSRC, Denmark;
IN2P3-CNRS and CEA-DRF/IRFU, France;
SRNSFG, Georgia;
BMBF, HGF and MPG, Germany;
GSRI, Greece;
RGC and Hong Kong SAR, China;
ISF and Benoziyo Center, Israel;
INFN, Italy;
MEXT and JSPS, Japan;
CNRST, Morocco;
NWO, Netherlands;
RCN, Norway;
MEiN, Poland;
FCT, Portugal;
MNE/IFA, Romania;
MESTD, Serbia;
MSSR, Slovakia;
ARRS and MIZ\v{S}, Slovenia;
DSI/NRF, South Africa;
MICINN, Spain;
SRC and Wallenberg Foundation, Sweden;
SERI, SNSF and Cantons of Bern and Geneva, Switzerland;
MOST, Taiwan;
TENMAK, T\"urkiye;
STFC, United Kingdom;
DOE and NSF, United States of America.
In addition, individual groups and members have received support from
BCKDF, CANARIE, Compute Canada and CRC, Canada;
PRIMUS 21/SCI/017 and UNCE SCI/013, Czech Republic;
COST, ERC, ERDF, Horizon 2020 and Marie Sk{\l}odowska-Curie Actions, European Union;
Investissements d'Avenir Labex, Investissements d'Avenir Idex and ANR, France;
DFG and AvH Foundation, Germany;
Herakleitos, Thales and Aristeia programmes co-financed by EU-ESF and the Greek NSRF, Greece;
BSF-NSF and MINERVA, Israel;
Norwegian Financial Mechanism 2014-2021, Norway;
NCN and NAWA, Poland;
La Caixa Banking Foundation, CERCA Programme Generalitat de Catalunya and PROMETEO and GenT Programmes Generalitat Valenciana, Spain;
G\"{o}ran Gustafssons Stiftelse, Sweden;
The Royal Society and Leverhulme Trust, United Kingdom.
 
The crucial computing support from all WLCG partners is acknowledged gratefully, in particular from CERN, the ATLAS Tier-1 facilities at TRIUMF (Canada), NDGF (Denmark, Norway, Sweden), CC-IN2P3 (France), KIT/GridKA (Germany), INFN-CNAF (Italy), NL-T1 (Netherlands), PIC (Spain), ASGC (Taiwan), RAL (UK) and BNL (USA), the Tier-2 facilities worldwide and large non-WLCG resource providers. Major contributors of computing resources are listed in Ref.~\cite{ATL-SOFT-PUB-2021-003}.

% End of text imported from the .//acknowledgements/Acknowledgements.tex input file

\printbibliography
 
% ATLAS Collaboration author list
% Reference date of EXOT-2022-37 is 2023-04-06
% Author list last updated on date 04-JAN-24
% Data extracted on 04-Jan-2024 for paper reference EXOT-2022-37
% at 11:35am
 
\begin{flushleft}
\hypersetup{urlcolor=black}
{\Large The ATLAS Collaboration}

\bigskip

\AtlasOrcid[0000-0002-6665-4934]{G.~Aad}$^\textrm{\scriptsize 102}$,
\AtlasOrcid[0000-0002-5888-2734]{B.~Abbott}$^\textrm{\scriptsize 120}$,
\AtlasOrcid[0000-0002-1002-1652]{K.~Abeling}$^\textrm{\scriptsize 55}$,
\AtlasOrcid[0000-0001-5763-2760]{N.J.~Abicht}$^\textrm{\scriptsize 49}$,
\AtlasOrcid[0000-0002-8496-9294]{S.H.~Abidi}$^\textrm{\scriptsize 29}$,
\AtlasOrcid[0000-0002-9987-2292]{A.~Aboulhorma}$^\textrm{\scriptsize 35e}$,
\AtlasOrcid[0000-0001-5329-6640]{H.~Abramowicz}$^\textrm{\scriptsize 151}$,
\AtlasOrcid[0000-0002-1599-2896]{H.~Abreu}$^\textrm{\scriptsize 150}$,
\AtlasOrcid[0000-0003-0403-3697]{Y.~Abulaiti}$^\textrm{\scriptsize 117}$,
\AtlasOrcid[0000-0003-0762-7204]{A.C.~Abusleme~Hoffman}$^\textrm{\scriptsize 137a}$
% ...
\bigskip
\\

% ...
$^{29}$Physics Department, Brookhaven National Laboratory, Upton NY; United States of America.\\
% ...
$^{35}$$^{(a)}$Facult\'e des Sciences Ain Chock, R\'eseau Universitaire de Physique des Hautes Energies - Universit\'e Hassan II, Casablanca;$^{(b)}$Facult\'{e} des Sciences, Universit\'{e} Ibn-Tofail, K\'{e}nitra;$^{(c)}$Facult\'e des Sciences Semlalia, Universit\'e Cadi Ayyad, LPHEA-Marrakech;$^{(d)}$LPMR, Facult\'e des Sciences, Universit\'e Mohamed Premier, Oujda;$^{(e)}$Facult\'e des sciences, Universit\'e Mohammed V, Rabat;$^{(f)}$Institute of Applied Physics, Mohammed VI Polytechnic University, Ben Guerir; Morocco.\\
% ...
$^{49}$Fakult\"{a}t Physik , Technische Universit{\"a}t Dortmund, Dortmund; Germany.\\
% ...
$^{55}$II. Physikalisches Institut, Georg-August-Universit\"{a}t G\"ottingen, G\"ottingen; Germany.\\
% ...
$^{102}$CPPM, Aix-Marseille Universit\'e, CNRS/IN2P3, Marseille; France.\\
% ...
$^{117}$Department of Physics, New York University, New York NY; United States of America.\\
% ...
$^{120}$Homer L. Dodge Department of Physics and Astronomy, University of Oklahoma, Norman OK; United States of America.\\
% ...
$^{137}$$^{(a)}$Departamento de F\'isica, Pontificia Universidad Cat\'olica de Chile, Santiago;$^{(b)}$Millennium Institute for Subatomic physics at high energy frontier (SAPHIR), Santiago;$^{(c)}$Instituto de Investigaci\'on Multidisciplinario en Ciencia y Tecnolog\'ia, y Departamento de F\'isica, Universidad de La Serena;$^{(d)}$Universidad Andres Bello, Department of Physics, Santiago;$^{(e)}$Instituto de Alta Investigaci\'on, Universidad de Tarapac\'a, Arica;$^{(f)}$Departamento de F\'isica, Universidad T\'ecnica Federico Santa Mar\'ia, Valpara\'iso; Chile.\\
% ...
$^{150}$Department of Physics, Technion, Israel Institute of Technology, Haifa; Israel.\\
$^{151}$Raymond and Beverly Sackler School of Physics and Astronomy, Tel Aviv University, Tel Aviv; Israel.\\
% ...
$^{a}$ Also Affiliated with an institute covered by a cooperation agreement with CERN.\\
$^{b}$ Also at An-Najah National University, Nablus; Palestine.\\
$^{c}$ Also at Borough of Manhattan Community College, City University of New York, New York NY; United States of America.\\
$^{d}$ Also at Center for High Energy Physics, Peking University; China.\\
$^{e}$ Also at Center for Interdisciplinary Research and Innovation (CIRI-AUTH), Thessaloniki; Greece.\\
$^{f}$ Also at Centro Studi e Ricerche Enrico Fermi; Italy.\\
% ...

\end{flushleft}

% Created with Glance <Atlas.Glance@cern.ch>

\end{document}